\documentstyle [epsfig,prl,aps]{revtex}
\begin{document}
\draft
\title{ SYMMETRIES, QUANTUM GEOMETRY, 
AND THE FUNDAMENTAL INTERACTIONS}
\large
\author{Jeeva Anandan}
\address{ Department of Physics and Astronomy, University of 
South
Carolina,\\ Columbia, SC 29208 \\E-mail: jeeva@sc.edu}
\date{December 1, 2000, revised March 23, 2001}
\maketitle
\bigskip
\begin{abstract}
\large
A generalized Noether's theorem and the operational 
determination of a physical geometry in quantum physics are used to motivate a
quantum 
geometry consisting of relations between quantum states that 
are 
defined by a universal group. Making these relations dynamical 
implies the non local effect of the fundamental interactions 
on the wave 
function, as in the Aharonov-Bohm 
effect and its generalizations to non Abelian gauge fields 
and gravity. The usual space-time geometry 
is obtained as the classical limit of this quantum geometry 
using the quantum state space metric.
\end{abstract}
\normalsize
\bigskip
\bigskip
\bigskip

\noindent{\it Paper presented at the XXIII International Colloquium on
Group Theoretical Methods in Physics, Dubna, Russia, 30 July 
- 05 August, 2000, quant-ph/0012011}.

\newpage
\section{Introduction}

The space-time geometry that is commonly used today arose from 
classical physics. An interesting question is what geometry 
is appropriate for quantum physics. It was suggested that the 
universal symmetry group elements which act on all Hilbert spaces may 
be appropriate for constructing a physical geometry for quantum theory 
\cite{an1980a}. I also proposed the systematic study of 
all the fundamental interactions {\it operationally} from their 
effects on quantum interference \cite{an1979}. The purpose 
of this paper is to attempt to bring together these two 
approaches. The 
modular variables introduced by Aharonov, Pendleton and 
Petersen \cite{ah1969} will play a useful role in this.

In section 2, I shall review  Noether's theorem and its 
converse in a generalized form in which the conserved 
quantities are elements of a group and not the generators of 
this group as usually stated. This will suggest a quantum 
geometry by relations defined by the universal group elements, 
which constitute 
the symmetry group of physics, that act on all Hilbert spaces, 
as discussed in section 3. 
The classical limit of this geometry will be obtained in section
4 as the usual space-time geometry from the quantum 
state metric in Hilbert spaces and the universality of the 
action of the translational group in every Hilbert space. 
In section 5, the non locality of 
fundamental interactions in quantum physics implied by 
this  approach, as shown 
physically by the Aharonov-Bohm effect \cite{ah1959} and 
its generalizations \cite{ah1967,an1979}, will be studied. The study of the 
gravitational Aharonov-Bohm effect around a cosmic string in particular 
suggests that the use of universal group elements as quantum 
distances may be appropriate.

\section{Some Reflections on Noether's Theorem}

The usual statement of Noether's theorem is that for every 
continuous
symmetry of the equations of motion (determined by the 
Lagrangian
or Hamiltonian) there exists a conserved quantity. 
Although this is easy to prove, the meaning of this theorem 
is more 
readily evident in the Hamiltonian than in the 
Lagrangian formulation. A symmetry of the equations of motion 
or 
time evolution is a transformation $s$ such that in any 
experiment if 
$s$ is applied to the initial state of the physical objects 
(fields, 
particles, etc.) participating in the experiment then the 
final state of 
the transformed experiment must be the same as $s$ applied to 
the 
final state of the original experiment. For example, spatial 
translational symmetry implies that if the 
apparatus is translated to a new spatial location then the 
same experiment should give the 
same result.

Suppose $U$ is the time-evolution operator, and $\psi_i$, 
$\psi_f$ 
are the initial and final states, i.e. $\psi_f=U\psi_i$. Then 
the above 
definition of $s$ being a symmetry of the time evolution is
$$s\psi_f=Us\psi_i$$
for every initial state $\psi_i$. This is equivalent to
the commutator
\begin{equation}
[ U,s ]=0.
\label{symmetry}
\end{equation}
But (\ref{symmetry}) states also that $s$ is conserved 
during the 
time-evolution. Therefore, the statements that $s$ is a 
symmetry and 
that $s$ is conserved are the {\it same} statement (\ref{symmetry}), 
and there is nothing to prove!

Now suppose that there is a continuous symmetry generated by 
$Q$. 
Then (\ref{symmetry}) is satisfied with $s=\exp(iQq)$ for 
all $q$ , and therefore 
\begin{equation}
[ U,Q]=0.
\label{commutator}
\end{equation}
Hence, $Q$ is conserved, which is Noether's theorem. 
Furthermore, if the Hamiltonian $H$ is independent of time 
$t$, as it is for an 
isolated system, then $U=\exp(-{i\over\hbar}Ht)$. If 
(\ref{commutator}) is valid 
for all $t$, then
\begin{equation}
[H,Q]=0.
\label{hcommutator}
\end{equation}
The above results may be extended to classical physics by 
turning the above 
commutators into Poisson brackets in classical phase space. 
These classical 
results may be regarded as the classical limit of the quantum 
results by 
recognizing that the symplectic structure that gives the 
Poisson brackets are the 
classical limit \cite{an1990b} of a symplectic structure in 
quantum theory 
\cite{ki1979} that gives the commutators.

But the 
conservation of $s$ that follows from (\ref{symmetry}) is 
more 
general than the  usual form of Noether's theorem. There are 
at least two 
situations in which (\ref{symmetry}) is valid but there are 
no corresponding 
(\ref{commutator}) or (\ref{hcommutator}). First, as is well known,
in both 
classical and quantum 
physics, $s$ may be a discrete symmetry 
instead of a continuous symmetry. For example, $s$ may be 
parity, 
which is a symmetry and therefore conserved for all 
interactions 
except the weak interaction, as far as we know. Another 
example is that 
(\ref{symmetry}) is satisfied for $s=\exp(iQq_k)$ for a 
discrete set of values 
$q_k$ only. Second, in quantum physics the mean value of 
$s=\exp(iQq)$ has more 
information than the mean value of all the moments of $Q$, 
namely $Q^n$ where $n$ 
is any positive integer \cite{ah1969}. This is unlike in classical
physics  where the mean value 
of a transformation generated by $Q$ may be obtained using 
the mean values of all 
the $Q^n$. Both these situations will be considered in 
section 5. 

Since in (\ref{symmetry}) $U$ and $s$ occur symmetrically, it 
follows that the 
converse of the generalized Noether's theorem is also true: {\it A 
transformation $s$ 
that is conserved must be a symmetry of the equations of 
motion.} The usual view is 
that $U$ is more fundamental than $s$ because $U$ is 
determined by the dynamical 
laws which are regarded as primary, whereas the symmetries 
such as $s$ are 
obtained secondarily as the symmetries of these laws. But the 
concise form 
(\ref{symmetry})  of the connection between the dynamical 
laws and symmetries, 
in which $U$ and $s$ are on an equivalent footing, suggest 
that we may equally 
turn the usual view around and regard the transformations $\{ s\}$ as 
primary and $U$ as 
derived from them to satisfy (\ref{symmetry}) so that $\{ s\}$  
are the symmetries \cite{an1999}. The possibility 
of regarding 
symmetries as fundamental relations between quantum states by 
associating them with 
a quantum geometry will be explored in the next section.

\section{Quantum Geometry}

The concept of space originates from our common experience of 
translating objects and from the possible states they can 
occupy. If 
we translate a cup, for example, in various possible ways, 
classically we may 
say that the different configurations or states of the cup 
are 
``immersed'' in ``space''.  This space is {\it universal} in 
the sense that 
it is regarded as independent of the objects which are 
``contained'' in 
it. 

But quantum mechanically it is not clear what is meant by the 
cup being 
``immersed'' in space. The cup consists of electrons,
protons and neutrons (or the quarks and gluons which make up the protons 
and 
neutrons), and the states of these particles 
belong to the corresponding Hilbert spaces which are different 
from the physical 
space or the phase space of classical physics. The 
translation of a cup 
therefore needs to be represented by the corresponding 
translation 
operators that act on these Hilbert spaces.  The fact that 
all the 
particles constituting the cup move together in some 
approximate 
sense suggests the introduction of universal translation 
group 
elements that are represented by operators that 
act on 
each Hilbert space. It is this {\it universality} of 
the 
translation group that gives us the concept of ``space'' that 
is 
independent of the particular system that partakes in it.

Also, it is well known that we cannot operationally determine 
the metric in space-
time, or even the points of space-time, using quantum probes 
\cite{wi1967} because 
of the uncertainty principle. If one tries to obtain the 
space-time geometry using 
a clock and radar light signals, which is possible in classical physics 
\cite{sy1960},
the uncertainty in the measurement 
of time in quantum physics is $\sim 
\hbar/\Delta E$, where $\Delta E$ is the uncertainty in the 
energy of the  clock. 
If we try to decrease this uncertainty by increasing $\Delta 
E$, then $\Delta 
E$ causes a corresponding uncertainty in 
the geometry of space-
time. The total uncertainty in the measurement of space-time 
distances is then 
$\hbar c/\Delta E + 2 G\Delta E/c^4$. The minimum value of 
this uncertainty as 
$\Delta E$ is varied is $\sim$ Planck-length 
$=\sqrt{G\hbar/c^3}$.  Hence, space-
time geometry is only approximately valid in quantum theory, 
with an uncertainty of the order of Planck length.

However, a geometry for quantum theory may be defined by 
relations determined by a 
universal group $S$, which generalizes the above translation 
group. This is 
universal in the sense that the same $S$ has a representation in each 
Hilbert space. But $S$ 
may have subgroups which may have trivial representations in 
some Hilbert spaces 
and not in others. An object may be displaced by any 
$s\epsilon S$, which means 
that $s$ acts on each of the Hilbert spaces of the particles or 
fields constituting that 
object through the corresponding representation of $S$. Each 
$\psi$ in each of 
these Hilbert spaces is mapped to a corresponding $\psi_s$ by 
this action of $s$, 
and the group element $s$ that determines 
the {\it relation} between $\psi$ and $\psi_s$ is 
independent of the Hilbert space and is therefore 
universal.
In the example of a cup 
considered above, $s$ is 
an element of the translation group, $\psi$ and $\psi_s$ are 
the states of each 
particle constituting the cup before and after the 
translation, and the relation 
between each such pair is universal in the sense that the 
entire cup has undergone 
this translation, or any other object that could take the place of the
cup. This quantum geometry cannot be subject to  the above criticisms 
of the space-time geometry because the action of $S$ on each 
Hilbert space is not 
subject to any uncertainty.

It is reasonable to require that this geometry is preserved 
in time in the absence 
of interactions. Then each $s\epsilon S$ is conserved and the 
converse of 
Noether's theorem stated in the last paragraph of the 
previous section implies 
that
$s$ is also a symmetry of the time evolution. Since the 
evolution equations are 
now determined by the standard model, $S$ may be the 
symmetry group of the 
present day standard model, namely $P\times U(1)\times 
SU(2)\times SU(3)$, where 
$P$ is the Poincare group. But if the standard model is 
superseded by new physics 
that has a different symmetry group then $S$ should 
be this new group and 
the above statements would all be unaffected.

As an illustration of the geometrical relations proposed 
here, consider the 
experimentally known quantization of electric charge:  
all known charges are 
integral multiple of the fundamental charge $e_0$. An aspect 
of this is that the 
magnitudes of the charges of the electron and the proton are 
experimentally known 
to be equal to an amazing precision.
To obtain charge quantization, take $s$ above to be an 
arbitrary element of the 
electromagnetic $U(1)$ group, which is a subgroup of $S$. 
This universal $U(1)$ 
group is a circle parametrized by $\Lambda$, say, that varies 
from $0$ to 
$\Lambda_0$ so that $0$ and $\Lambda_0$ represent the same 
point on this group, 
chosen to be the identity. Since $U(1)$ is abelian, it has 
only one-dimensional 
representations. Hence the action of $s(\Lambda)$ on an 
arbitrary state gives 
\begin{equation}
\psi_s= \exp(iQ\Lambda) \psi, 
\label{U1}
\end{equation}
where $Q$ corresponds to the particular representation of 
$U(1)$ in the Hilbert 
space in which $\psi$ belongs to. But since 
$s(\Lambda_0)=s(0)$, which is due to 
the compactness of the $U(1)$ group, $\exp(iQ\Lambda_0)=1$ 
for all 
representations \cite{ya1970}. Hence, $Q\Lambda_0=2\pi n$ or 
\begin{equation}
Q=n {2\pi \over\Lambda_0},
\label{quantization} 
\end{equation}
where $n$ is an integer.

To interpret $Q$, consider the physical implementation of the 
transformation $s$. 
This may be done by sending each of the particles through the 
same electromagnetic 
field with 4-vector potential $A_\mu$ in a particular gauge 
so that the effect of 
the electromagnetic field alone on the particle is given by
\begin{equation}
\psi_s= \exp( -i{q\over \hbar c} \int A_\mu (x) dx^\mu) \psi,
\label{em} 
\end{equation}
which is a $U(1)$ transformation. Indeed, the statement that 
the electromagnetic 
field is a $U(1)$ gauge field may be taken to mean that it is 
physically possible 
to implement a $U(1)$ gauge transformation using the 
electromagnetic field in this 
way. Then $q$ has the interpretation of the electric charge. 
Comparing (\ref{em}) 
with (\ref{U1}), we may take $\Lambda$ to be $\int A_\mu 
dx^\mu$ in which case 
$Q={q\over\hbar c}$. Hence, from (\ref{quantization}),
\begin{equation}
q=n e_0
\label{quant}
\end{equation}
where $e_0={2\pi \hbar c\over\Lambda_0}$ is a universal 
constant that is 
determined experimentally to be
${1\over 3} e$, where $e$ is the charge of the electron. The 
exact equality of the 
magnitudes of the charges of the electron and the proton may 
now be understood as 
due to them belonging to representations corresponding to 
$n=3$ and $n=-3$, 
respectively.

The above argument also provides a reason for the 
introduction of Planck's 
constant which is purely geometrical. The exponent in 
(\ref{em}) must be dimensionless because the expansion of the 
exponential has all 
powers of the exponent. Now, $ {q\over c} \int A_\mu dx^\mu$ 
is meaningful in 
classical physics. But to turn it into a dimensionless 
quantity, it is necessary 
to introduce a new scale, which is provided by $\hbar$. From 
the present point of 
view, this is needed in order to physically implement the $U(1)$ group 
elements that define 
relations between states which are part of the quantum 
geometry. Also, from 
(\ref{quant}), $q$ is proportional to $e$, and $A_\mu$ is also 
proportional 
to $e$ because the charges that generate $A_\mu$ via 
Maxwell's equations are 
proportional to $e$. Hence, the exponent
in (\ref{em}) is 
proportional to the fine-structure constant $e^2\over \hbar 
c$. This argument may 
be extended to gauge fields in general, and dimensionless 
coupling constants are 
obtained for all of them. From now onwards, units in which $c=1$ 
will be used, and the metric
convention is $(+,-,-,-)$.

The relation defined by (\ref{em}) is not gauge invariant. 
Hence, it cannot be 
used to define an invariant geometrical `distance'. Consider 
again the translation  
of a cup which may be performed by acting on all the quantum 
states of the 
particles constituting the cup by a universal group element $ 
\exp(-{i\over\hbar}\hat p\ell) $, where $\hat p$ is a 
generator of translation. 
The action of this group element on a wave function is also not gauge
invariant. But we may  combine the two transformations 
to define the gauge-covariant transformation 
$\psi_{\mbox{\boldmath $\ell$}}(x)=f_{\mbox{\boldmath $\ell$}} (x)\psi(x)$, 
where $x$ stands for $x^\mu$ or equivalently $({\bf x}, t)$, and \cite{fourier}
\begin{equation}
f_{\mbox{\boldmath $\ell$}}(x) = \exp (-{i\over\hbar}\hat{\bf p}\cdot {\mbox{\boldmath $\ell$}}) 
\exp \{i{q\over \hbar} \int_{\bf x}^{{\bf x}+{\mbox{\boldmath $\ell$}}} {\bf A}({\bf y},t)\cdot
d{\bf y}\}
\label{p}
\end{equation}
where at present the integral is taken along the straight line 
joining $x=({\bf x},t)$ and $({\bf x}+{\mbox{\boldmath $\ell$}},t)$ for simplicity. Then, clearly,
\begin{equation}
f_{\mbox{\boldmath $\ell$}}(x) =  \exp \{i{q\over \hbar} 
\int_{{\bf x}-{\mbox{\boldmath $\ell$}}}^{{\bf x}} {\bf A}({\bf
y},t)\cdot d{\bf y}\} \exp (-{i\over\hbar}\hat{\bf p}\cdot {\mbox{\boldmath $\ell$}}) .
\label{reversed}
\end{equation}
It is easy to show, using (\ref{p}) 
and (\ref{reversed}), that the set of operators 
$\{f_{\mbox{\boldmath $\ell$}}|\mbox{\boldmath $\ell$}\epsilon {\cal R}^3\}$ 
is a group under multiplication.

From (\ref{p}), under a gauge
transformation, $\psi'(x)=  u(x)\psi(x)$, where $u(x)= \exp\{i{q\over \hbar} 
\Lambda(x)\}$, and
$ A_\mu'(x) = A_\mu(x)-\partial_\mu \Lambda (x)$, $f_{\mbox{\boldmath $\ell$}}$ transforms to
$$f'_{\mbox{\boldmath $\ell$}}(x)=\exp (-{i\over\hbar}\hat{\bf p}\cdot {\mbox{\boldmath $\ell$}})
~ u({\bf x}+{\mbox{\boldmath $\ell$}},t) 
\exp \{ i{q\over \hbar} \int_{\bf x}^{{\bf x}+{\mbox{\boldmath $\ell$}}} {\bf A}({\bf y},t)\cdot
d{\bf y}\}  u({\bf x},t)$$
On using 
\begin{equation}
\exp (-{i\over\hbar}\hat{\bf p}\cdot {\mbox{\boldmath $\ell$}}) u({\bf x}+{\mbox{\boldmath $\ell$}},t)=u({\bf x},t) \exp
(-{i\over\hbar}\hat{\bf p}\cdot {\mbox{\boldmath $\ell$}})
\label{p'}
\end{equation}
it follows that
\begin{equation}
 f'_{\mbox{\boldmath $\ell$}}(x)= u({\bf x},t) \exp (-{i\over\hbar}\hat{\bf p}\cdot {\mbox{\boldmath $\ell$}}) 
\exp \{i{q\over \hbar} \int_{\bf x}^{{\bf x}+{\mbox{\boldmath $\ell$}}} {\bf A}({\bf y},t)\cdot
d{\bf y} \} u^\dagger({\bf x},t)=u(x)f_{\mbox{\boldmath $\ell$}}(x)u^\dagger(x)
\end{equation}
Hence, $ f_{\mbox{\boldmath $\ell$}}$ acts gauge covariantly on the Hilbert space. 

$ f_{\mbox{\boldmath $\ell$}}$ may also shown to be gauge covariant from the fact that
\begin{eqnarray}
<\psi|f_{\mbox{\boldmath $\ell$}}|\psi>&=&<\exp ({i\over\hbar}{\bf p}\cdot {\mbox{\boldmath $\ell$}})\psi|
\exp \{i{q\over \hbar} \int_{\bf x}^{{\bf x}+{\mbox{\boldmath $\ell$}}} {\bf A}({\bf y},t)\cdot
d{\bf y} \}|\psi>\nonumber \\&& =\int d^3x \psi^\dagger({\bf x}+{\mbox{\boldmath $\ell$}},t)
\exp \{i{q\over \hbar} \int_{\bf x}^{{\bf x}+{\mbox{\boldmath $\ell$}}} {\bf A}({\bf y},t)\cdot
d{\bf y}\} \psi({\bf x},t)
\label{<f>}
\end{eqnarray}
is gauge invariant because the integrand is gauge invariant \cite{an1986}.
It also follows from (\ref{<f>}) that the operator (\ref{p}) is 
observable. For 
example, in the Josephson effect, where the current depends 
on the gauge invariant 
phase difference across the junction, if $\mbox{\boldmath $\ell$}$ is chosen 
to be the 
vector across the junction then 
$f_{\mbox{\boldmath $\ell$}}$ is observable 
from the current \cite{an1986}. 

A gauge may be chosen, even if the field strength is non vanishing, so that the component of $\bf A$ in the direction of 
$\mbox{\boldmath $\ell$}$ is zero. Then in this gauge $f_{\mbox{\boldmath $\ell$}}
=\exp (-{i\over\hbar}\hat{\bf p}\cdot {\mbox{\boldmath $\ell$}})$, which is conserved 
for an isolated system. Since $f_{\mbox{\boldmath $\ell$}}$ 
is gauge covariant, it follows that (\ref{p}) is conserved for an isolated system in every gauge.
$f_{\mbox{\boldmath $\ell$}}$ is a gauge 
covariant generalization of the modular momentum \cite{ah1969}, and may be called the 
modular kinetic momentum. When two systems interact, there would be an exchange of modular 
kinetic momentum. Indeed, this may be regarded as the definition of two systems interacting \cite{an1999}. 
As will be
seen in section \ref{interactions}, this exchange may happen even when there 
are no forces between the two systems, which makes the latter interaction more general than 
the usual interaction via forces. If
$\tilde\psi$ is the momentum space wave function of
$\psi$, then clearly
\begin{equation}
<\psi|\exp (-{i\over\hbar}\hat{\bf p}\cdot {\mbox{\boldmath $\ell$}})|\psi>
=\int d^3p~|\tilde\psi({\bf p},t)|^2~\exp (-{i\over\hbar}{\bf p}\cdot {\mbox{\boldmath $\ell$}})
\label{ft}
\end{equation}
Hence, the change in $<\psi|\exp (-{i\over\hbar}\hat{\bf p}\cdot {\mbox{\boldmath $\ell$}})|\psi>$ 
results in a change in
the distribution of the probability density $|\tilde\psi({\bf p},t)|^2$ of momentum $\bf p$ \cite{ah}. 
Since in the above gauge the gauge invariant
$<\psi|f_{\mbox{\boldmath
$\ell$}}|\psi>$ is the same as (\ref{ft}), it follows that, in an arbitrary gauge, an interaction results in a change in
the distribution of the gauge invariant kinetic momentum ${\bf p}-q{\bf A}$.

Similarly, the modular energy \cite{ah1969} may be generalized gauge-covariantly 
as follows. Define the transformation on the Hilbert space $f_\tau$ by $\psi_\tau 
(x)=f_\tau (x)\psi(x)$, where
\begin{equation}
f_\tau(x) = \{T\exp (-{i\over\hbar}\int_t^{t+\tau} H~dt)\}^\dagger ~\exp \{-i{q\over \hbar} 
\int_t^{t+\tau} A_0(x) dt\}
\label{H}
\end{equation}
and $T$ denotes time-ordering. To prove that $f_\tau$ is gauge-covariant under the 
above gauge transformation, note first that for a solution $\psi({\bf x}, t)$ of 
Schr\"odinger's equation, 
\begin{equation}
\psi( {\bf x}, t+\tau ) =T\exp(-{i\over\hbar}\int_t^{t+\tau}H~dt) \psi({\bf
x},  t).
\label{se}
\end{equation}   
Therefore, under the gauge transformation $\psi'( {\bf x}, t )=u({\bf x}, t)\psi( {\bf x},
t ),$
$$\psi'( {\bf x}, t+\tau )\equiv u({\bf x}, t+\tau)\psi( {\bf x}, t+\tau ) = u({\bf x}, t+\tau)
T\exp(- {i\over\hbar}\int_t^{t+\tau}H~dt) u^\dagger ({\bf x}, t) \psi'({\bf x}, t).$$ 
using (\ref{se}). Hence,
$$u({\bf x}, t+\tau) T\exp(-{i\over\hbar}\int_t^{t+\tau}H~dt) u^\dagger ({\bf x}, t)
= T\exp(-{i\over\hbar}\int_t^{t+\tau}H'~dt),$$ 
where $H'$ is the gauge transformed
Hamiltonian. The last equation is equivalent to
\begin{equation}
\{T\exp(-{i\over\hbar}\int_t^{t+\tau}H'~dt)\}^\dagger ~u({\bf x}, t+\tau) =
u({\bf x}, t)\{T\exp(-{i\over\hbar}\int_t^{t+\tau}H~dt)\}^\dagger
\label{H'}
\end{equation}
Under this gauge transformation, $f_\tau (x)$ 
transforms to 
\begin{eqnarray}
f_\tau '(x)&\equiv& \{ T\exp (-{i\over\hbar}\int_t^{t+\tau} H'~dt)\}^\dagger ~\exp \{-i{q\over
\hbar} 
\int_t^{t+\tau} A_0'(x) dt\}\nonumber \\&& =\{ T\exp (-{i\over\hbar}\int_t^{t+\tau}
H'~dt)\}^\dagger ~u({\bf x},  t+\tau) \exp \{-i{q\over \hbar} \int_t^{t+\tau} A_0(x) dt\}
u^\dagger({\bf x},  t)\nonumber \\&& = u({\bf x}, t)f_\tau u^\dagger({\bf x}, t)
\end{eqnarray}
on using (\ref{H'}). Hence, $f_\tau$ acts gauge-covariantly on the Hilbert space.

More generally, for an arbitrary gauge field the above arguments hold 
with $eA_\mu$ replaced by $g_0 A_\mu^k T_k$ where $T_k$ generate the gauge group,
and $u(x)$ is the corresponding local gauge transformation. The gauge field
exponential (parallel transport) operator will be generalized to be along an arbitrary
piece-wise differentiable curve $\hat\gamma$ in ${\cal R}^4$. Path ordering is necessary because
$\{T_k\}$ do not commute in general. The energy-momentum operator $\hat p_\mu$ is defined by
$\hat p_0=H, \hat p_i =i\hbar {\partial\over \partial x^i}$ . Now define the transformation
$g_\gamma$ on the Hilbert space by
$\psi_\gamma (x)=g_\gamma(x)\psi(x)$, where
\begin{equation}
g_\gamma(x)= P\exp(-{i\over\hbar}\int_{\bar\gamma} \hat p_\mu dy^\mu)~P \exp \{-i{g_0\over
\hbar}
\int_\gamma A_\mu^k (y)T_k 
dy^\mu\}
\label{g}
\end{equation}
with P denoting path ordering, and $\gamma$ is a curve in space-time that is congruent to $\hat\gamma$ while $\bar\gamma$ is the curve
$\gamma$ traversed in the reverse order. In (\ref{g}), $\gamma$ begins at $x$ and ends at $x+\ell$,
where $\ell^\mu$ is a fixed vector (independent of $x^\mu$),  and $\bar\gamma$ therefore begins at $x+\ell$
and ends in $x$. Then
\begin{equation}
P\exp(-{i\over\hbar}\int_{\bar\gamma} \hat p_\mu dy^\mu)=\{ P\exp(-{i\over\hbar}\int_{\gamma}
\hat p_\mu dy^\mu)\}^\dagger
\label{conjugate}
\end{equation}

Under a local gauge transformation, (\ref{g}) transforms to 
\begin{eqnarray}
g_\gamma '(x) &\equiv& P\exp(-{i\over\hbar}\int_{\bar\gamma} \hat p_\mu ' dy^\mu) ~P\exp
\{-i{g_0\over
\hbar}
\int_\gamma {A'}_\mu^k (y)T_k dy^\mu\}
\nonumber \\&& =P\exp(-{i\over\hbar}\int_{\bar\gamma} \hat p_\mu ' dy^\mu)~u(x+\ell) ~P\exp
\{-i{g_0\over\hbar}
\int_\gamma A_\mu^k (y^\mu)T_k dy^\mu\}~u^\dagger(x)
\label{g'}
\end{eqnarray}
where ${\hat p'}_0=H',~~ {\hat p'}_i =\hat p_i=i\hbar {\partial\over \partial x^i}$. Now write 
$P\exp(-{i\over\hbar}\int_{\bar\gamma} \hat p_\mu ' dy^\mu)$ as a product of infinitesimal exponentials of
the form $(1-{i\over\hbar}p_j dy^j)(1-{i\over\hbar}p_0' dy^0)$, and use (\ref{p'}) and (\ref{H'}) in their
infinitesimal forms. Then (\ref{g'}) implies
\begin{eqnarray}
g_\gamma '(x) &=& u(x)~P\exp(-{i\over\hbar}\int_{\bar\gamma} \hat p_\mu  dy^\mu) ~P\exp
\{-i{g_0\over\hbar}
\int_\gamma A_\mu^k (y)T_k dy^\mu\}~u^\dagger(x)
\nonumber \\&&=u(x)g_\gamma (x) u^\dagger (x)
\end{eqnarray}
Hence, the operator $g_\gamma$ acts gauge covariantly on the Hilbert space. It follows from this proof
that  $g_\gamma$ would be gauge-covariant also when $\bar\gamma$ in (\ref{g}) is replaced by any
piece-wise differentiable curve that would connect
$x+\ell$ to $x$. 

Also, on using (\ref{conjugate}),
\begin{eqnarray}
<\psi|g_\gamma|\psi>&=&<P\exp(-{i\over\hbar}\int_{\gamma}
\hat p_\mu dy^\mu)\psi|P\exp(-
ig_0 \int_\gamma 
A_\mu^k T_k ~dy^\mu)|\psi>\nonumber \\&& =\int d^3x ~{\psi_\gamma}^\dagger(x+\ell)P\exp(-ig_0 
\int_\gamma 
A_\mu^k T_k ~dy^\mu)\psi(x)
\label{expectation}
\end{eqnarray}
where $\psi_\gamma(x+\ell)\equiv P\exp(-{i\over\hbar}\int_{\gamma}
\hat p_\mu dy^\mu)\psi(x)$. The expectation value (\ref{expectation}) is gauge invariant because the
integrand is gauge invariant. 
It may be
observable, in principle, by the Josephson effect for a non Abelian 
gauge theory
proposed in \cite{an1986}. More experimental consequences of $g_\gamma$ will be discussed in section V.

The energy-momentum operator $\hat p_\mu$ may be generalized to relativistic quantum field theory
by defining its components as the conserved quantities obtained via Noether's theorem from the invariance 
of the Lagrangian under space-time translations:
\begin{equation}
\hat p_\mu =\int d^3x~\hat T_{\mu 0}
\end{equation}
where  $\hat T_{\mu \nu}$ is the conserved, normal-ordered energy-momentum tensor.
Then the canonical commutation relations imply
that $\hat p_\mu$ generates space-time translations. It therefore follows from arguments analogous to the
above that
$g_\gamma$, given by (\ref{g}) with the quantum field theoretic $\hat p_\mu$, is gauge covariant.
Since $\hat p_\mu$ transforms as a covariant vector under Lorentz transformations, it also follows that
$g_\gamma$ is Lorentz invariant. Moreover, $g_\gamma$ may be generalized to curves $\gamma$ in ${\cal
R}^n$ ($n$ is any positive integer), which would correspond to an $n-$dimensional space-time.

It may be reasonable to take
$g_\gamma$ as a quantum  distance that replaces the classical space-time distance along the curve
$\gamma$. But to do so it would be necessary to obtain the classical distance in an appropriate limit from
$g_\gamma$, which will be studied in the next section.

\section{Classical Limit}

To take the classical limit of this geometry, note that 
classical space-time is constructed with measuring 
instruments consisting of particles that have approximate 
position and momentum. It is therefore reasonable to 
represent them by Gaussian wave packets which have minimum 
uncertainty. For a particle with mean position at the origin 
and mean momentum zero, the normalized wave function of such a 
state up to an arbitrary phase factor is
\begin{equation}
\psi_{\bf 0}({\bf x}) = (2\pi \Delta x^2)^{-1/4}
\exp(-{{\bf x}^2\over 4\Delta x^2})
\label{gaussian}
\end{equation}
where $\Delta x$ is the uncertainty in position. This may be a 
state of a molecule in the cup, mentioned in section 3, in a harmonic oscillator 
potential in which case it would not 
spread. As the cup is displaced, the above wave function 
becomes
\begin{equation}
\psi_{\mbox{\boldmath $\ell$}}({\bf x})\equiv \exp(-{i\over \hbar}\hat{\bf p}\cdot {\mbox{\boldmath $\ell$}}) 
\psi_{\bf 0}({\bf x})= (2\pi \Delta x^2)^{-1/4}
\exp(-{({\bf x}-{\mbox{\boldmath $\ell$}})^2\over 4\Delta x^2})
\label{translation}
\end{equation}
up to a phase factor.

In (\ref{p}) the second factor may be made the identity by 
choosing a gauge in which ${\bf A}\cdot {\mbox{\boldmath $\ell$}} =0$. Therefore,  $\exp(-i\hat{\bf 
p}\cdot {\bf 
\ell})$ may be regarded as a special case of (\ref{p}) or (\ref{g}), and hence as defining a 
quantum
distance between the states $\psi_{\bf 0}$ and $\psi_{\mbox{\boldmath $\ell$}}$. We may 
therefore expect
a metric to be defined on  the translation group to which this operator belongs to 
and 
use that to define a metric in space. But this 
group, being Abelian, has no natural metric on it. There are 
two ways, however, that a metric may be defined on it. One is to 
use the Casimir operator of the Poincare group, 
$\eta_{ab}P^aP^b$, to define a metric on it, which locally 
may be associated with the space-time metric \cite{an1980a}. 
The other method, which will be used here, is to utilize the 
overlap of the two wave functions to obtain a
measure of the displacement between them, which would then 
give an equivalent metric in the translation group. This is possible if 
the space of wavefunctions on which this group acts has an inner 
product, which would give 
a measure of the overlap and therefore how far a state has been 
translated. 

Such a measure is given by
the Fubini-Study metric in the quantum state 
space, or the set of rays, of every Hilbert space.  This is 
the unique metric, up to multiplication by an overall 
constant, that is invariant under unitary (and anti-unitary) 
transformations. This may therefore be written in the form 
\cite{pr1980,an1990a,an1990b}
\begin{equation}
dS^2 = 4(1-| <\psi|\psi'> |^2)
\label{fs}
\end{equation}
where $dS$ is the infinitesimal distance between two 
neighboring states (rays) represented by normalized state 
vectors $\psi$ and $\psi'$. Clearly, $dS$ is zero when the 
states are the same, and it increases when the overlap 
between the states decreases. It is also invariant under 
unitary transformations, and must therefore be the Fubini-
Study metric. The factor $4$ in (\ref{fs}) is just a 
convention which ensures that this metric in the state space 
of the Hilbert space spanned by $\psi$ and $\psi'$ 
is the metric on a sphere of unit radius.

Now substitute $\psi_{\mbox{\boldmath $\ell$}}({\bf x})$ and 
$\psi_{\mbox{\boldmath $\ell$}} +d{\mbox{\boldmath $\ell$}}({\bf x}) $
as $\psi$ and $\psi'$ in 
(\ref{fs}). Then,
\begin{equation}
dS^2 = { d{\mbox{\boldmath $\ell$}}^2\over \Delta x^2}
\label{metric}
\end{equation}
neglecting higher order terms in $ d{\mbox{\boldmath $\ell$}}$ because it is 
infinitesimal.
Hence $d{\mbox{\boldmath $\ell$}}^2$, which is the same for all Hilbert spaces,
may be used as a metric on the $3$ 
dimensional translational group that is parametrized by the components of
the vector
${\mbox{\boldmath $\ell$}}$. Locally, this metric may be regarded as a metric in 
the physical space of classical physics.

This result may be generalized to an arbitrary state $\psi$ as follows. 
Require that $<\psi|\psi'>$ to be close to the identity, where 
\begin{equation}
\psi'({\bf x})\equiv
\exp(-{i\over \hbar}\hat{\bf p}\cdot d{\mbox{\boldmath $\ell$}})
\psi ({\bf x}). 
\label{spatial}
\end{equation}  
Therefore expand
\begin{equation}
<\psi|\psi'>\simeq <\psi|(1-{i\over \hbar}\hat{p} d\ell -{1\over 
2\hbar^2}\hat{p}^2 d\ell^2)\psi>=1-{i\over
\hbar} <\psi|p|\psi> d\ell -{1\over 2\hbar^2}<\psi|p^2|\psi>  d\ell^2
\end{equation}
where $\hat{p}$ is the momentum component in the direction of $d{\mbox{\boldmath $\ell$}}$. 
Substituting this in
(\ref{fs}),
\begin{equation}
dS^2 = {4\Delta p^2\over\hbar^2} d{\mbox{\boldmath $\ell$}}^2 
\label{metric2}
\end{equation} 
where $\Delta p$ is the uncertainty in $p$: $\Delta p^2=<\psi|p^2|\psi>-
<\psi|p|\psi>^2$. For the Gaussian wave packet, $\Delta p ~\Delta x=\hbar/2$ and 
therefore (\ref{metric2})  gives (\ref{metric}) in this case. 

Time is measured by a clock. Since the clock must have moving 
parts, the uncertainty $\Delta E$ of its Hamiltonian $\hat{H}$ must 
be non zero. Neglecting any external interaction of the 
clock, $\hat{H}$ is a constant. The infinitesimal time evolution  of the the 
clock is given by
\begin{equation}
|\psi(t+dt)> = \exp(-{i\over \hbar}\hat{H} dt) |\psi(t)>
\label{evolution}
\end{equation}
The Fubini-Study distance $dS$ along the evolution 
curve in the quantum 
state space corresponding to $|\psi(t)>$ is obtained analogous to the
derivation of (\ref{metric2}) to be
\cite{an1990a,an1990b} 
\begin{equation}
dS^2={4\Delta E^2\over\hbar^2}~ dt^2
\label{clock}
\end{equation}
A quantum clock directly measures the Fubini-Study distance 
$S$ and the time $t$ is then inferred from $S$ using 
(\ref{clock}).
The appearance of the same $t$ in (\ref{clock})
in all Hilbert spaces 
is due to the universality of the action of the time 
translation $\exp({i\over \hbar}\hat{H} dt)$ in every Hilbert 
space. This is analogous to the universality of the spatial 
displacements which was used earlier to obtain the spatial metric. 

The spatial translation (\ref{spatial}) and the 
active time translation corresponding to (\ref{evolution}) 
may be written
covariantly as 
\begin{equation}
\psi'=  \exp({i\over\hbar} \hat{p}_\mu d\ell^\mu)\psi_o
\label{covariant}
\end{equation}
where $d\ell^\mu=(cdt, d{\mbox{\boldmath $\ell$}})$. This transformation is the infinitesimal 
version of
(\ref{g}) in a special gauge that makes the second factor in (\ref{g}) the identity \cite{inf}. 
In an arbitrary gauge,
the  same space and
time metrics are obtained by replacing the operator in (\ref{covariant}) with 
the infinitesimal version of (\ref{g}) in the above
treatment. In relativistic quantum theory, owing to the transformation property 
of
$\hat{p}_\mu$ mentioned at the end of section III, these metrics give a space-time metric that is
invariant under Lorentz transformations.  In the presence of gravity, 
this Lorentzian metric may be obtained locally
by doing the above space- time
translations in a freely falling Einstein elevator, which 
then globally gives a curved pseudo-Riemannian
metric.

\section{Interactions and the Non Locality of Quantum Theory}
\label{interactions}

If the 
quantum geometry is determined by relations 
between states that are group elements, and if these group 
elements, which are our observables,
are made dynamical the way Einstein made space-time distances 
dynamical in order to obtain gravity, 
then this would 
give both gravity and gauge fields \cite{an1999}. Also, quantum 
mechanics has an 
inherent non locality, which may also be understood as due to 
these group elements being the basic observables. The 
combination of these two statements imply that gauge fields 
and gravity should affect quantum states in a non local 
manner as in the Aharonov-Bohm effect \cite{ah1959}, as 
will be discussed later. 

First consider the non locality of quantum theory, which 
may be illustrated by the following 
example: Electrons 
with initial momentum in the $x-$ direction go 
through an infinite diffraction grating 
in the $yz-$plane of a Cartesian coordinate system, with the 
length of the slits 
along the $z$-direction. Then the grating destroys continuous 
translational symmetry for the electrons in the $y-$ direction. 
However, if 
the 
distance between successive slits in the $y$ direction is 
$\ell$ then 
$s= \exp(-i {p \ell\over \hbar})$ satisfies (\ref{symmetry}), 
where $p$ is the 
momentum operator for electrons in the $y-$direction which 
generates 
translations in the $y-$direction. (For simplicity, here and 
henceforth, the $\hat{}$ over operators is omitted.) 
Hence, it follows from the  
generalized Noether's theorem in section
2 that
$\exp(-i{p 
\ell\over \hbar})$ is 
conserved although $p$ is {\it not} conserved. Indeed, it is 
well known 
that the interference 
fringes on a screen that is parallel to and far away from the 
$yz$-
plane is given by $\ell \sin\theta_n =n\lambda$, where 
$\lambda$ is the wave length and $n$ is an 
integer. Therefore, the possible values of the momentum for an 
electron in 
the $y-$ direction after the interaction are 
$p_n={h\over \lambda} 
\sin\theta_n=n{h\over \ell}$, i.e. $ \exp(-i {p_n \ell\over 
\hbar})=1$. Hence,
$\exp(-i {p \ell\over \hbar})$ is conserved during the passage 
of electrons through the 
grating.

The above operator $s$ is 
equivalent to the modular 
momentum $p(mod {h\over \ell})$ introduced by Aharonov et al 
on the basis of the above example \cite{ah1969}. But 
here I shall treat $s$ as an element of a universal group 
that is used to define a 
quantum geometry as in section 3. $s$ may be obtained from 
experiments by measuring 
the Hermitian observables
\begin{equation}
s_R \equiv {1\over 2}[\exp(-i {p \ell\over \hbar})+ 
\exp(i{ p \ell\over \hbar})],
s_I \equiv {1\over 2i}[\exp(-i {p \ell\over \hbar})- \exp(i{ p 
\ell\over \hbar})].
\end{equation}
Therefore, the unitary operator $s$ may also be regarded as 
an observable. It is 
important to note that this is a {\it non-local} observable, 
unlike $p$. 

This non locality in quantum 
mechanics may be 
illustrated also in the simple interference experiment of two 
coherent wave packets.
Suppose that the two wave packets are moving in the $x-$direction 
and have no overlap at time $t$. For simplicity, assume that they
are the same except that their centers are 
separated by a displacement $\ell$ in the 
$y-$direction. Let $\alpha$ be the phase difference 
between the wave packets. The wave is then a 
superposition of these two wave packets: 
\begin{equation}
\psi(x,y,z,t)= {1\over\sqrt{2}}\{ \phi(x,y-\ell,z,t)+ e^{i\alpha} 
\phi(x,y,z,t)\}
\end{equation}
Now no local experiments performed on the two wave packets at 
the two slits could 
determine the phase factor $ e^{i\alpha} $. For example, the 
expectation  values 
of the local variables $p^n$, where $n$ is any positive 
integer, give no 
information about $ e^{i\alpha}$ \cite{ah1969}. This is easily verified by 
writing 
$p^n= (-i\hbar {\partial\over \partial y})^n$ in the 
coordinate representation. 
But
\begin{equation}
<\psi|\exp(-i {p \ell\over \hbar})|\psi>= {e^{i\alpha} \over 2} 
\label{alpha} 
\end{equation}
This means that the momentum distribution at time $t$ does 
depend on the phase 
factor $e^{i\alpha } $, i.e. if $p$ is measured then the 
probability distribution 
for obtaining the individual eigenvalues of $p$ is changed by 
this phase factor (but the average $<p>$ is unchanged). 
And this may be experimentally verified by letting the wave 
packets interfere and 
observing the shift in interference fringes. Hence, 
$<\psi|\exp(-i {p \ell\over 
\hbar})|\psi>$ contains more information than the expectation 
values 
$<\psi|p^n|\psi>$ of any of the moments of momentum $p^n$. This 
is basically due to the linear structure of the Hilbert space, 
which physically corresponds to the principle of superposition, 
and the fact that $\psi$ is not an analytic function.

This fundamental non locality of quantum mechanics translates 
into a non locality 
of the effect of all the fundamental interactions on the wave 
function. This has 
been shown for the Aharonov-Bohm effect due to 
a magnetic field by Aharonov et al \cite{ah1969}. It may be 
illustrated in the above described interference of two wave 
packets as follows. Suppose the above two wave packets $A$ and $B$ 
are those of an electron and they pass on the two sides of a 
solenoid containing a magnetic flux $\Phi$. The gauge may be 
chosen so that the vector potential is non zero only along a 
thin strip bounded by two planes indicated by the 
dotted lines in figure 1. 
Then when the line $AB$ 
passes the solenoid, the wave packet $A$ acquires 
a phase difference $\alpha = {e\over\hbar c}\Phi$ 
with respect to the
wave packet $B$. 
Therefore, the expectation value of the modular 
momentum $s$, given by (\ref{alpha}), 
which was $1/2$ before the line $AB$ passed the solenoid is 
now $\exp(i{e\over\hbar c}\Phi)/2$. It was pointed out by 
Aharonov \cite{ah} that in the last statement $s$
may be 
replaced by the gauge invariant modular kinetic momentum $f_{\mbox{\boldmath $\ell$}}$ given by
(\ref{p}). This is because before and after the wave 
packets pass the solenoid the vector potential $\bf A$ is zero along the line $AB$ and 
therefore $f_{\mbox{\boldmath $\ell$}}$ is the same as $s$. Since $f_{\boldmath \ell}$ is gauge 
covariant, $<\psi| f_{\mbox{\boldmath $\ell$}}|\psi>$ is gauge invariant. Hence, the 
above mentioned change in $<\psi| f_{\mbox{\boldmath $\ell$}}|\psi>$ by
the factor $\exp(i{e\over\hbar c}\Phi)$, 
as the line $AB$ crosses the solenoid, is the same in every gauge.

\centerline{\psfig{figure=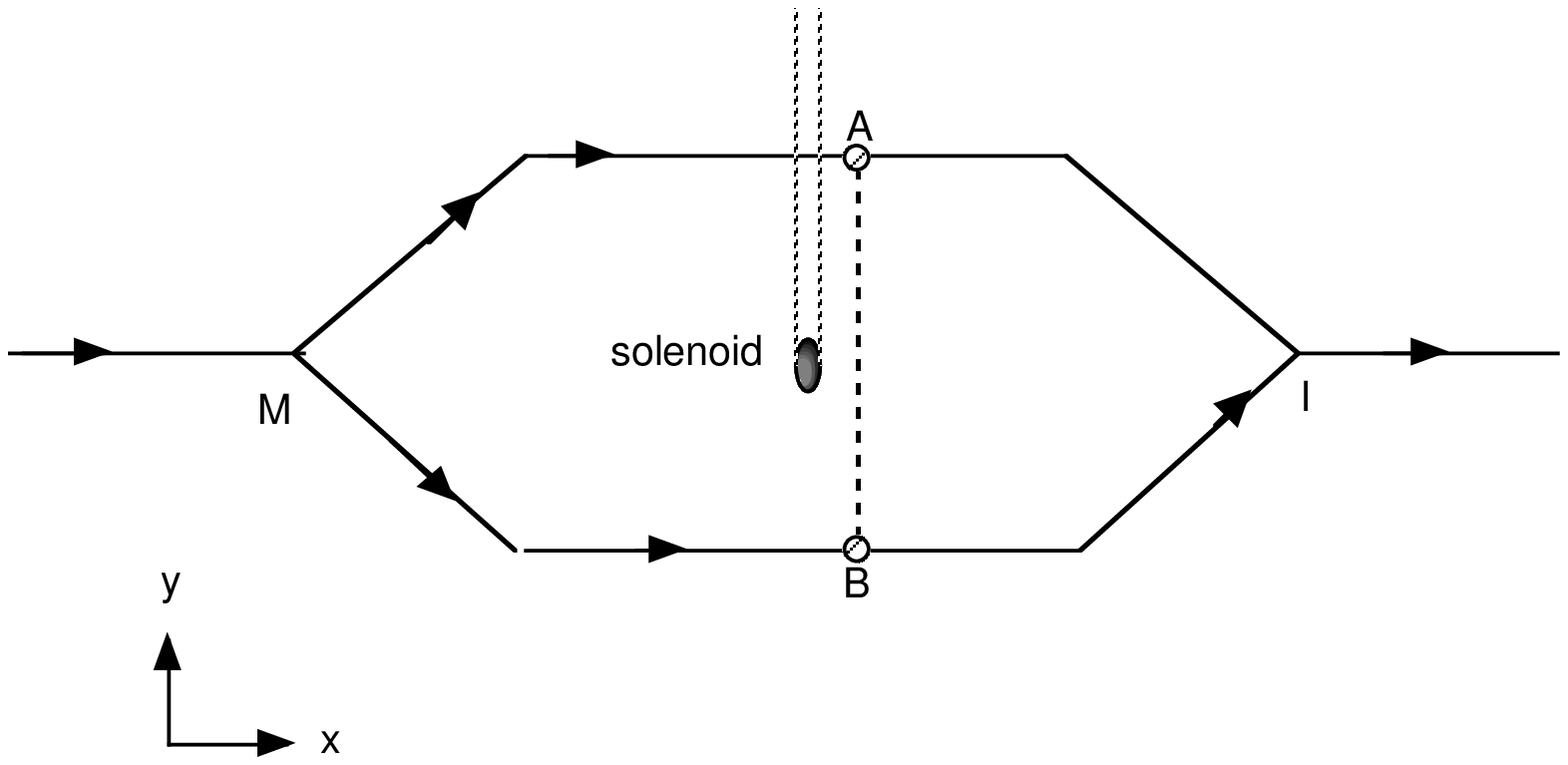,height=3.5in}}

\noindent\footnotesize
Figure 1. Vector Aharonov-Bohm effect in which a wave packet 
of an electron is 
split coherently into two wave  packets
at the beam splitter $M$ which are then made to interfere at 
$I$. When the imaginary line joining the centers of the
wave packets $A$ and $B$
sweeps across the 
magnetic flux in the solenoid (shaded region) 
the modular momentum or the modular kinetic momentum 
associated with this line changes, as pointed out by 
Aharonov \cite{ah}.
\bigskip
\normalsize

However, we may replace $ f_{\mbox{\boldmath $\ell$}}$ in the above arguments by the more 
general gauge covariant operator $g_\gamma$ given by (\ref{g}) with $\gamma$ here being an arbitrary
(piece-wise differentiable) space-like curve that joins $A$ and $B$. Then 
\begin{equation}
g_\gamma(x) = \exp (-{i\over\hbar}\hat{\bf p}\cdot {\mbox{\boldmath $\ell$}}) 
\exp \{i{e\over \hbar} 
\int\limits_{~\gamma~\bf x}^{{\bf x}+{\mbox{\boldmath $\ell$}}} {\bf A}({\bf y},t)\}\cdot
d{\bf y}
\end{equation}
where the integral is from $\bf x$ to ${\bf x}+{\mbox{\boldmath $\ell$}}$ along $\gamma$. Hence,
\begin{equation}
<\psi| g_\gamma |\psi> =\int d^3x \psi^\dagger({\bf x}+{\mbox{\boldmath $\ell$}},t)
\exp \{i{e\over \hbar} 
\int\limits_{~\gamma~\bf x}^{{\bf x}+{\mbox{\boldmath $\ell$}}} {\bf A}({\bf y},t)\cdot
d{\bf y}\} \psi({\bf x},t)
\label{gamma}
\end{equation}
which is gauge invariant. Then $<\psi| g_\gamma |\psi>$ 
changes by the factor 
$e^{\pm i{e\over\hbar c}\Phi}$  as a portion of $\gamma$ crosses the solenoid. 
It follows that there is nothing special about the straight line $AB$ 
crossing the solenoid in the experiment described in figure 1. 
The arbitrariness of the choice of $\gamma$
joining $A$ and $B$ in the above argument reflects the topological nature 
of the Aharonov-Bohm effect.
What is ultimately observed in this effect is the phase factor 
$\exp(-i{e\over \hbar} \oint A_\mu dx^\mu)$
where the integral is around the solenoid or more generally around the region 
in which the field strength is non
vanishing. Since this is an integral of the $1-$form $A_\mu$ along a curve, 
it contains no
information about the metric of space-time\cite{NR}. To determine the straight line, 
or a geodesic in general, the
metric is needed, and the above phase factor therefore cannot show any preference to a geodesic such as
the line $AB$. The integrand of (\ref{gamma}) (with $e$ replaced by 2$e$) 
was previously used in ref. \cite{an1986} to study the Josephson effect due to the enclosed magnetic flux in a
superconducting ring that has a Josephson junction, which is also an Aharonov-Bohm effect. 

Consider now the scalar Aharonov-Bohm effect. A wave packet 
traveling in the $x-$ direction is partially transmitted and 
reflected by a beam splitter $M$ (figure 2a). 
The resulting two wave packets 
which travel in opposite directions are reflected by two 
mirrors $M_1$ and $M_2$ situated along the 
$x-$axis and they interfere 
subsequently. Meanwhile a pair of oppositely charged 
capacitor plates $C$ is separated and closed so that there is a 
non zero electric field in the region enclosed by the world-
lines of the centers of the wave packets in the $xt-$plane. 
The same experiment 
is viewed in the rest frame of the reflected wave packet in 
figure 2b.

\centerline{\psfig{figure=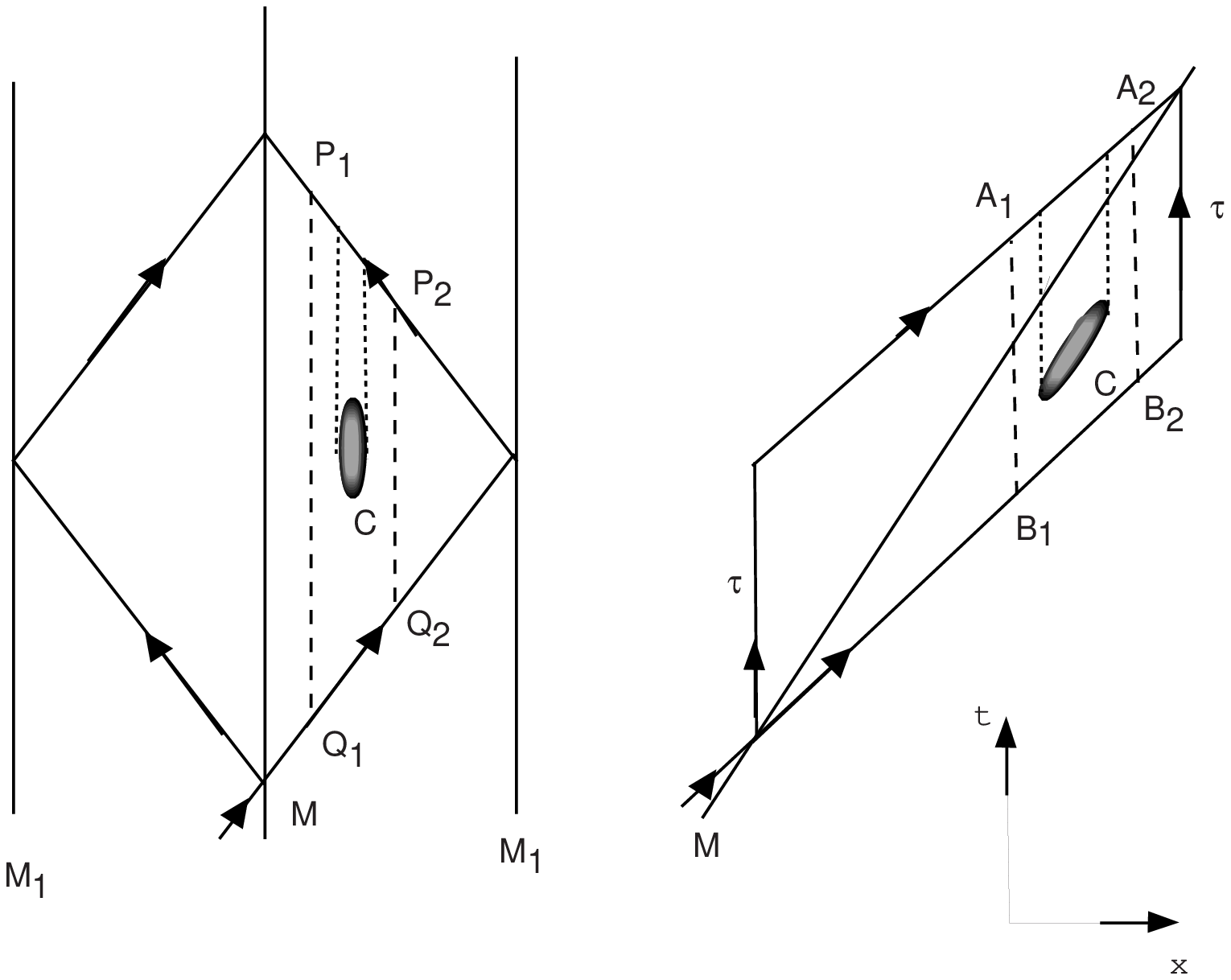,height=3.5in}}

\noindent\footnotesize
Figure 2. A scalar Aharonov-Bohm effect shown schematically in the $tx-$plane. 
(a) The modular energy and modular kinetic energy 
associated with the imaginary lines $P_1 Q_1$ and $P_2 Q_2$ are different, 
partly because of the scalar $AB$ phase shift due to the region of 
non zero electric field in the
capacitor that is open and shut (shaded region). (b) The same experiment 
viewed in the
rest frame of the wave packet that is reflected at the beam splitter $M$. 
When the imaginary line $AB$ 
joining the wave
packets  sweeps across the region of non zero electric field in the
capacitor the modular energy and modular kinetic energy 
associated with this line changes due to the AB phase shift.

\bigskip\normalsize

We may choose a gauge in which the vector 
potential is non zero only along a strip between the dotted 
lines parallel to the time-axis in figure 2a or 2b. Then the wave 
packet $A$ at time $t+\tau$ develops a phase shift $\beta$ with 
respect to $B$ at time $t$ as the imaginary line $AB$ crosses the 
space-time region containing the electric field, where
\begin{equation}
\beta={e\over \hbar}\int_C F_{0x}dx dt
\end{equation}
and $C$ is the region in which the electric field $E= F_{0x}$ 
is non zero. This is a non local effect which may be 
understood using $U=\{ T\exp(-{i\over 
\hbar}\int_0^\tau Hdt)\}^\dagger$, where $T$
denotes time ordering and $\tau$ is the time interval 
between the events $A_1$ and$B_1$. We shall assume that the time dependence 
of $H$ comes only from the vector potential contained in $H$. 
Suppose the wave function of the electron is
$$\psi={1\over\sqrt{2}}(\psi_A+\psi_B)$$ 
where
$\psi_A$ and
$\psi_B$ are the wave  functions of the above two localized wave packets. Then 
\begin{equation}
<\psi(t)|U|\psi(t)>= {1\over 2}<\psi_A(t)|U|\psi_B(t)>= {1\over 
2}<\psi_A(t+\tau)|\psi_B(t)>
\label{scab}
\end{equation}
changes by the factor $e^{i\beta}$ due to 
the electric field $E$ in the region $C$ as the imaginary line $AB$ sweeps across 
the 
small region $C$ where $E$ is non zero (fig. 2b). 

In the last statement and in (\ref{scab}) we 
may replace $U$ by the gauge covariant modular kinetic energy 
operator $f_\tau$ given by (\ref{H}), 
with $e$ replacing $q$. This is because $A_0$ is zero along $A_1B_1$ 
and $A_2B_2$, which are outside the strip that contains the non zero $A_\mu$. 
However, 
\mbox{$<\psi(t)|f_\tau|\psi(t)>$} is gauge invariant and therefore its change 
mentioned above may be obtained in any gauge. But
since there are no forces  acting on the electron, 
there is no change of its kinetic energy $
H-eA_0$ or  any 
of its moments. Hence, $<\psi(t)|f_\tau|\psi(t)>$ contains non local aspects which 
cannot be obtained by measurements of the kinetic energy operator or any of its 
moments.
Thus the  scalar Aharonov-Bohm effect may be viewed as a quantum effect which
is {\it non local in time}.

The above results are easily generalized to the Aharonov-Bohm effects 
due to non Abelian gauge
fields\cite{an1979} by  replacing 
the electromagnetic fluxes by fluxes of Yang-Mills field strength, and using the expectation 
values
(\ref{expectation}) of the gauge covariant operator $g_\gamma$ given by (\ref{g}). 
In all cases, the gauge invariant $<\psi|g_\gamma|\psi>$ changes as $\gamma$ 
passes a cross-section of the gauge field
flux. If the gauge field flux has a singular cross-section, then $\gamma$ passes 
the flux at an event, which of course is
independent of the inertial frame in which the effect is being described. 
When this happens, in this idealized case, $<\psi|g_\gamma|\psi>$ changes to 
$<\psi|g_\gamma ~P \exp \{-i{g_0\over
\hbar}
\oint_C A_\mu^k (y)T_k 
dy^\mu\}|\psi>$, where $C$ is a closed curve going around the Yang-Mills flux.

The group element (\ref{g}) belongs to the group $T_4\times G$, 
where for a closed system $T_4$ is the translation group and $G$ is the gauge 
group. 
But it has an asymmetry
in that the part of (\ref{g}) that belongs to $G$ is 
dynamical, whereas the part that belongs to $T_4$ is fixed. 
Since I proposed that the fundamental interactions should 
correspond to the universal group element (\ref{g}) being 
dynamical, consistency requires that the part of (\ref{g}) that 
belongs to $T_4$ should be dynamical as well, i.e. $\ell^\mu$ 
should be made dynamical. But the classical space-time geometry was 
constructed in section 4 using the latter group elements. 
It follows therefore that making $\ell^\mu$ dynamical would make 
the space-time metric dynamical and not fixed as it is in Minkowski space-time. 
Therefore, the interaction that corresponds to making the $T_4$ group elements 
dynamical gives, in the classical limit, the well known geometrical description of 
gravity in 
classical general relativity. A generalization of it is obtained by replacing 
$T_4$ with $T_n$, where $n$ is any positive integer. 
Thus the present approach requires the existence of gravity.

I now give a simple illustration of the above unified 
way of treating gravity and gauge fields by 
considering the gravitational analog of the above vector 
Aharonov-Bohm effect. The geometry surrounding a 
cosmic string in the two dimensional section normal to the 
axis of the string at a given time is that of a cone whose 
center is at the axis, which is seen by solving the 
classical gravitational field equations \cite{an1996}. 
The space-time geometry of a non-
rotating cosmic string is obtained by simply adding to this 
plane the extra dimension in the direction of the axis and 
the time dimension; then the curvature outside the string 
is zero everywhere (figure 3). 
It is known that this
geometry is similar to the  electromagnetic field
around a solenoid because the curvature is zero outside the string and yet there 
is a non
trivial holonomy around it. 

Consider now two wave packets separated by a distance
$\ell$ and whose centers move along initially 
parallel lines  such that the geodesic line 
$AB$ joining their centers meets the
conical  singularity $S$ at its midpoint. But there 
is another geodesic that connects the same pair of points $A$ 
and $B$ of length $\ell \cos({\theta\over 2})$, shown by the 
line $A'B$ in figure 3, where $A'$ is identified with $A$. (Actually, there are 
two geodesics connecting $A$ and $B$ when the angle $A\hat SB$ on the left side 
exceeds $\pi-\theta$ but is less than $\pi$. The corresponding angles $A\hat SB$ 
on the right side are $\pi$ and $\pi-\theta$, respectively. The 
\bigskip
\centerline{\psfig{figure=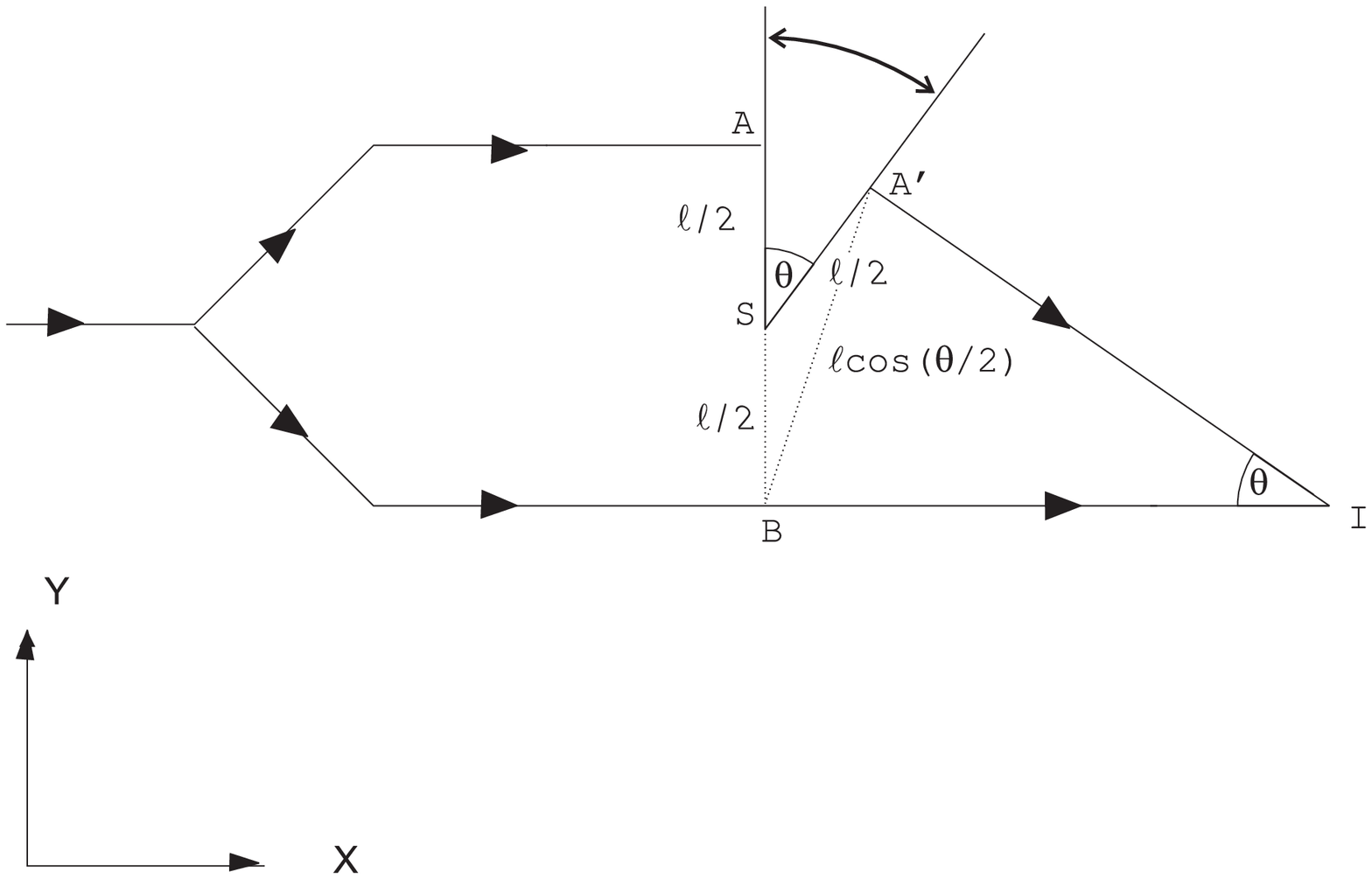,height=3.2in}}

\noindent\footnotesize
Figure 3. A gravitational analog of the experiment in fig. 1. 
The conical geometry around a cosmic string that is normal to the plane 
through $S$ is represented by
cutting off the wedge $ASA'$ from flat space and identifying the planes along  
which  it is
cut. The wave packets moving at $A$ (same as $A'$) and $B$ are focused by this 
geometry to
interfere at $I$. Just before $AB$ crosses $S$, there are two geodesics 
connecting 
$A$ and $B$ 
of lengths $\ell$ and $\ell\cos(\theta/2)$ in the curvature free region. 
But just after the crossing there is a 
unique
geodesic $A'B$ of length $\ell\cos(\theta/2)$ joining $A$ and $B$.
\bigskip\normalsize
\linebreak
singularity may be 
replaced by a small smooth cap that merges with the rest of the cone smoothly. Then there 
would be a third geodesic joining $A$ and $B$ though this cap.)
Just after $AB$ crosses the conical singularity, there is 
only one geodesic joining $A$ and $B$, whose length 
is $\ell\cos(\theta/2)$. This is somewhat analogous to the 
change in modular kinetic 
momentum when $AB$ crosses the solenoid in fig. 
1 , because
$\exp(-i{\bf p}\cdot{\mbox{\boldmath $\ell$}})$ now translates locally along parallel geodesic 
line segments of
length $\ell$, and this length has gotten shorter after the crossing.

In figure 4, this 
result is generalized to the case of $S$ not 
being the midpoint of the geodesic $ASB$, and it is also 
seen to  be `gauge' independent in the sense of being 
invariant under the rotation of the `wedge' mentioned above. There is then a phase 
shift $\Delta \phi$ due to the 
difference in path lengths traveled by the wave packets given 
by 
\begin{equation}
\Delta \phi= {p_0\over \hbar} (d_1-d_2) \simeq {p_0\over 2\hbar} 
(\ell_2-\ell_1)\theta==4\pi G\mu {p_0\over \hbar} (\ell_2-
\ell_1)
\end{equation}
for small $\theta$, where and $d_1$ and $d_2$ are the path lengths $AI$ and $BI$, 
$\mu$ is the mass per unit length 
of the cosmic string, and $p_0$ is the initial momentum of the 
particle. If the particle 
carries spin, then there is also a phase 
shift due to the coupling of spin to the curvature. 

\bigskip

\centerline{\psfig{figure=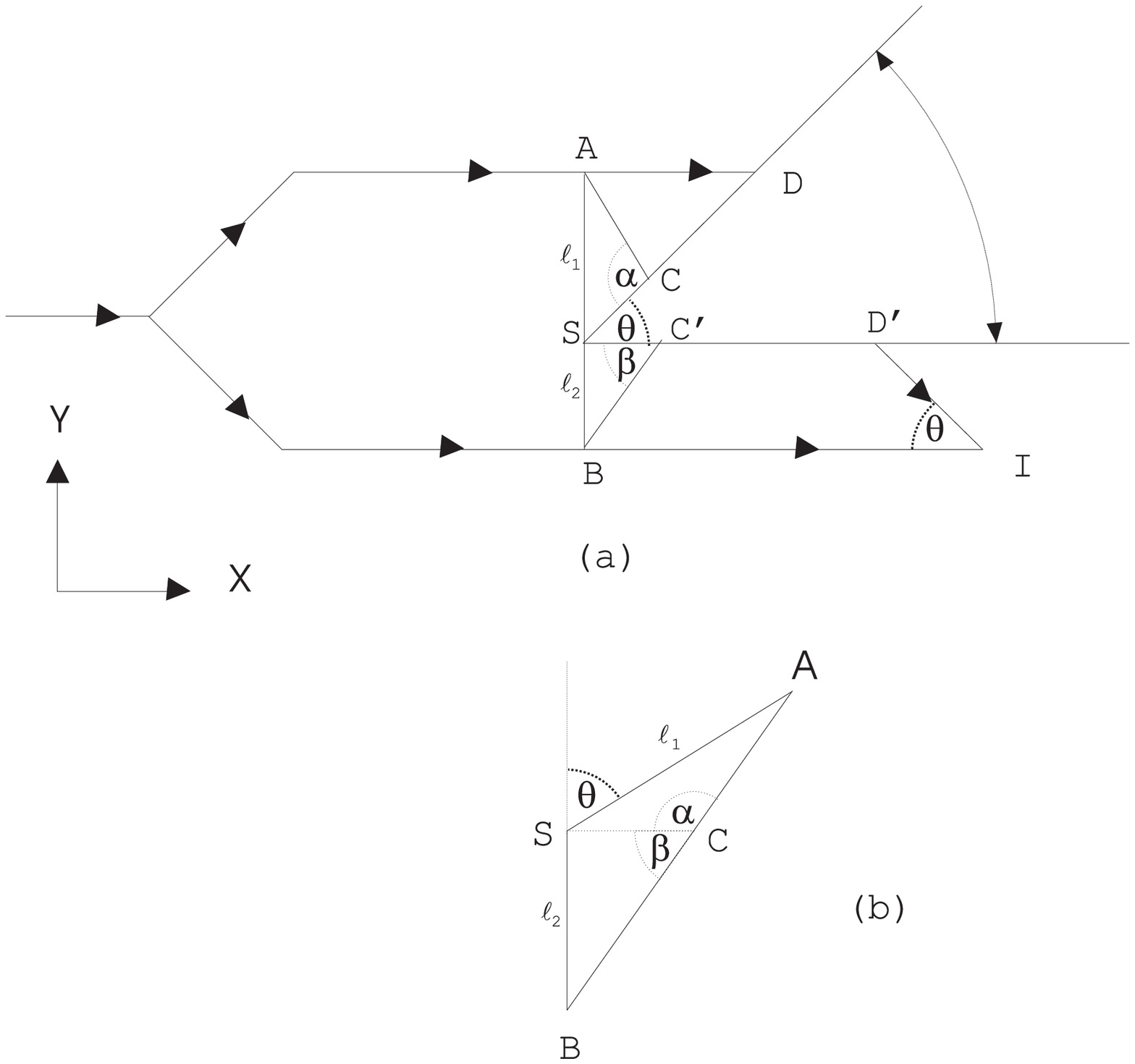,height=3.3in}}

\noindent\footnotesize
Figure 4. A generalization of the gedanken experiment shown 
schematically in fig. 3 to unequal lengths $AS=\ell_1$ and $BS=\ell_2$. The wedge 
$DSD'$ has an arbitrary orientation. But
since $SC$ and $SC'$ are identified, and $\alpha+\beta=\pi$ in order for $ACB$ to 
be a geodesic, the distances along the geodesics $ASB$ and 
$ACB$ are respectively $\ell_1+\ell_2$ and 
$(\ell_1^2+\ell_2^2+2\ell_1\ell_2\cos\theta)^{1/2}$, as shown in fig. (b), 
independently of the orientation of the wedge. 
\bigskip

\normalsize
This and 
other phase shifts for interference of two wave packets 
around a cosmic string are studied elsewhere and are 
understood as being due to the Poincare holonomy around the 
string \cite{an1994,an1996}. In the present approach, these phase shifts 
may be understood by associating with each curve an element of the Poincare
group element \cite{an1999} whose expectation value changes as the curve crosses the cosmic string.

\section{DISCUSSION AND CONCLUSION}

The change in geodesic distances between the wave 
packets due to the cosmic string, in section 5, is not surprising because gravity 
changes distances, according to general relativity, and the 
cosmic string is a purely general relativistic object without 
a Newtonian analog. What may be more interesting is the similar 
change of  the `quantum distances' due to the 
electromagnetic field in the usual Aharonov-Bohm effect and its 
generalization to non Abelian gauge fields. Both these effects 
may be treated in a somewhat analogous manner
if the modular kinetic energy-momentum (\ref{g}), regarded here as universal group elements, may
be  interpreted as `distances' in a quantum geometry as proposed earlier. The fact 
that space-time distances may be obtained from them approximately, as shown in 
section 4, reinforces this view. 

The treatment of (\ref{g}) as an observable, which implies the above
mentioned non local effects in quantum theory, perhaps removes the mystery of 
why
although the interactions are  local as they occur in  the Hamiltonian or
Lagrangian, nevertheless there are non- local effects such as 
the Aharonov-Bohm and its generalizations to non 
abelian gauge fields and 
gravitation \cite{ya1983}.

A criticism that may be made against regarding the universal group elements 
(\ref{g}) as quantum distances is that their action on wave functions seems to 
require that $\hat\gamma$ be interpreted as a set of space-time 
curves, whereas it was argued in section 3 that space-time geometry 
is not appropriate for quantum theory. However, as discussed 
in that section, this limitation becomes critical only at the Planck scales. 
But at the Planck scale these group elements need not be associated with curves 
in space-time. They may be defined simply as operators acting on quantum states 
defining the quantum geometry and representing the fundamental interactions.  Thus 
it is possible to have a quantum geometry even at scales in which the space-time 
geometry breaks down and this would also give a quantum description of all the 
interactions.

\bigskip

\noindent
ACKNOWLEDGMENTS
\bigskip

I thank Yakir Aharonov, Alonso Botero and Ralph Howard for useful discussions. This work was partially supported by NSF
grant 9971005 and ONR grant N00014-00-1-0383.


\begin{thebibliography}{99}

\bibitem{an1980a}
J. Anandan, Foundations of Physics {\bf 10,} 601-629 (1980),

\bibitem{an1979}
J. Anandan, Il Nuovo Cimento {\bf 53A.}, 221 (1979).

\bibitem{ah1969}
Y. Aharonov, H. Pendleton and A. Peterson, Int. J. 
Theoretical Phys. {\bf 2,} 213 (1969).

\bibitem{ah1959}
Y. Aharonov and D. Bohm, Phys. Rev. {\bf 115,} 485 (1959).

\bibitem{ah1967}
D. Wisnievesky and Y. Aharonov, Ann. Phys. {\bf 45,} 479 (1967).

\bibitem{an1990b}
J. Anandan, 
Phys. Lett. {A 147,} 3 (1990); J. Anandan, Foundations of 
Physics, {\bf 21,} 1265 (1991).


\bibitem{ki1979}
T. W. B. Kibble, Communications in Mathematical Physics, {\bf 
65,} 189 (1979).

\bibitem{an1999}
J. Anandan, Foundations of Physics, {\bf 29,} no. 11, 1647-
1672 (1999), quant-ph/9808045.

\bibitem{wi1967}
Eugene Wigner, {\it Symmetries and Reflections} 
(Indiana U. P., Bloomington/London, 1967)

\bibitem{sy1960}
J. L. Synge, {\it Relativity: The General Theory} (North Holland, Amsterdam, 
1960), Chapter III.

\bibitem{ya1970}
C.N. Yang, Phys. Rev. D, {\bf 1,} 2360 (1970).

\bibitem{fourier}
The action of $ \exp (-{i\over\hbar}\hat{\bf p}\cdot {\mbox{\boldmath $\ell$}})
=\exp (-{\mbox{\boldmath $\ell$}}\cdot \nabla)$ on an analytic wave function 
$\psi$ may be understood as the usual Taylor expansion of $\psi$. If $\psi$ 
is not analytic, then $\exp
(-{i\over\hbar}\hat{\bf p}\cdot {\mbox{\boldmath $\ell$}})$ acts on the momentum 
space wave function $\tilde\psi$ according
to $\tilde\psi({\bf p},t)\rightarrow \exp (-{i\over\hbar}{\bf p}\cdot {\mbox{\boldmath $\ell$}})\tilde\psi({\bf p},t).$ In
either case, we obtain $\exp (-{i\over\hbar}\hat{\bf p}\cdot {\mbox{\boldmath $\ell$}})\psi({\bf x},t)= \psi({\bf
x}-{\mbox{\boldmath $\ell$}},t)$.

\bibitem{an1986}
J. Anandan,  Phys. Rev. D, {\bf 33}, 2280-2287 (1986).

\bibitem{ah}
Y. Aharonov (private communication).

\bibitem{pr1980}
J.P. Provost and G. Vallee, Commun. Math. Phys. {bf 76,} 289 
(1980).

\bibitem{an1990a}
J. Anandan and Y. Aharonov, Physical Review Letters 65, 1697-1700 (1990),

\bibitem{inf}
If $d\ell^\mu$ is along $\gamma$ then $dy^\mu=-d\ell^\mu$ because $\bar\gamma$ 
is the reversal of $\gamma$.

\bibitem{NR}
This explains why in the non relativistic limit for the charged particle, 
the Aharonov-Bohm phase shift that it experiences due to the phase factor
$\exp(-i{e\over \hbar} \oint A_\mu dx^\mu)$ remains the same. This is because 
the latter phase factor is unaffected by the change in the space-time metric 
that takes place in the non relativistic limit.

\bibitem{an1996}
J. Anandan, Phys. 
Rev. D, {\bf 53,} 779 (1996), gr-qc/9507049, and references therein.

\bibitem{an1994}

J. Anandan, Physics Letters A, {\bf 195,} 284 (1994).

\bibitem{ya1983}
This reconciles the views of Y. Aharonov and C. N. Yang, who 
regarded the Aharonov-Bohm effect as being non local and local, 
respectively, in Proceedings of the International Symposium 
on the Foundations of Quantum Mechanics, Tokyo, August 1983, 
edited 
by S. Kamefuchi et al. (Physical Society of Japan, Tokyo, 
1984) 65-73.

 

\end{thebibliography}
\end{document}